\documentclass[twocolumn]{revtex4}

\usepackage{graphicx}

\newcommand{\beight}{\mbox{$^8$B}}

\newcommand{\cossun}{\mbox{$\cos\theta_{\rm sun}$}}
\newcommand{\thetasun}{\mbox{$\theta_{\rm sun}$}}
\newcommand{\thetaz}{\mbox{$\theta_z$}}
\newcommand{\nsixteen}{\mbox{$^{16}$N}}
\newcommand{\osixteen}{\mbox{$^{16}$O}}
\newcommand{\erecoil}{\mbox{$E_{\rm recoil}$}}
\newcommand{\cosz}{\mbox{$\cos \theta_z$}}

\newcommand{\hep}{\mbox{{\it hep}}}
\newcommand{\nhit}{\mbox{$N_{hit}$}}

\begin{document}


\title{Solar \beight\ and {\it hep} Neutrino Measurements from 1258 Days of 
Super-Kamiokande Data
}

\newcounter{foots}
\newcounter{notes}
\newcommand{\authoraticrr}{$^{1}$}
\newcommand{\authoratbu}{$^{2}$}
\newcommand{\authoratbnl}{$^{3}$}
\newcommand{\authoratuci}{$^{4}$}
\newcommand{\authoratcsu}{$^{5}$}
\newcommand{\authoratgmu}{$^{6}$}
\newcommand{\authoratgifu}{$^{7}$}
\newcommand{\authoratuh}{$^{8}$}
\newcommand{\authoratkek}{$^{9}$}
\newcommand{\authoratkobe}{$^{10}$}
\newcommand{\authoratkyoto}{$^{11}$}
\newcommand{\authoratlanl}{$^{12}$}
\newcommand{\authoratlsu}{$^{13}$}
\newcommand{\authoratlsuumd}{$^{13,14}$}
\newcommand{\authoratumd}{$^{14}$}
\newcommand{\authoratduluth}{$^{15}$}
\newcommand{\authoratsuny}{$^{16}$}
\newcommand{\authoratniigata}{$^{17}$}
\newcommand{\authoratosaka}{$^{18}$}
\newcommand{\authoratseoul}{$^{19}$}
\newcommand{\authoratshizuokasc}{$^{20}$}
\newcommand{\authoratshizuoka}{$^{21}$}
\newcommand{\authorattohoku}{$^{22}$}
\newcommand{\authorattokyo}{$^{23}$}
\newcommand{\authorattokai}{$^{24}$}
\newcommand{\authorattit}{$^{25}$}
\newcommand{\authoratwarsaw}{$^{26}$}
\newcommand{\authoratuw}{$^{27}$}

\newcommand{\addressoficrr}[1]{$^{1}$ #1 }
\newcommand{\addressofbu}[1]{$^{2}$ #1 }
\newcommand{\addressofbnl}[1]{$^{3}$ #1 }
\newcommand{\addressofuci}[1]{$^{4}$ #1 }
\newcommand{\addressofcsu}[1]{$^{5}$ #1 }
\newcommand{\addressofgmu}[1]{$^{6}$ #1 }
\newcommand{\addressofgifu}[1]{$^{7}$ #1 }
\newcommand{\addressofuh}[1]{$^{8}$ #1 }
\newcommand{\addressofkek}[1]{$^{9}$ #1 }
\newcommand{\addressofkobe}[1]{$^{10}$ #1 }
\newcommand{\addressofkyoto}[1]{$^{11}$ #1 }
\newcommand{\addressoflanl}[1]{$^{12}$ #1 }
\newcommand{\addressoflsu}[1]{$^{13}$ #1 }
\newcommand{\addressofumd}[1]{$^{14}$ #1 }
\newcommand{\addressofduluth}[1]{$^{15}$ #1 }
\newcommand{\addressofsuny}[1]{$^{16}$ #1 }
\newcommand{\addressofniigata}[1]{$^{17}$ #1 }
\newcommand{\addressofosaka}[1]{$^{18}$ #1 }
\newcommand{\addressofseoul}[1]{$^{19}$ #1 }
\newcommand{\addressofshizuokasc}[1]{$^{20}$ #1 }
\newcommand{\addressofshizuoka}[1]{$^{21}$ #1 }
\newcommand{\addressoftohoku}[1]{$^{22}$ #1 }
\newcommand{\addressoftokyo}[1]{$^{23}$ #1 }
\newcommand{\addressoftokai}[1]{$^{24}$ #1 }
\newcommand{\addressoftit}[1]{$^{25}$ #1 }
\newcommand{\addressofwarsaw}[1]{$^{26}$ #1 }
\newcommand{\addressofuw}[1]{$^{27}$ #1 }

\author{
{\large The Super-Kamiokande Collaboration} \\ 
\bigskip
S.~Fukuda\authoraticrr,
Y.~Fukuda\authoraticrr,
M.~Ishitsuka\authoraticrr, 
Y.~Itow\authoraticrr,
T.~Kajita\authoraticrr, 
J.~Kameda\authoraticrr, 
K.~Kaneyuki\authoraticrr,
K.~Kobayashi\authoraticrr, 
Y.~Koshio\authoraticrr, 
M.~Miura\authoraticrr
% ...
\\
\smallskip
\footnotesize
\it
\addressoficrr{Institute for Cosmic Ray Research, University of Tokyo, Kashiwa,Chiba 277-8582, Japan}\\
\addressofbu{Department of Physics, Boston University, Boston, MA 02215, USA}\\
\addressofbnl{Physics Department, Brookhaven National Laboratory, Upton, NY 11973, USA}\\
\addressofuci{Department of Physics and Astronomy, University of California, Irvine, Irvine, CA 92697-4575, USA }\\
\addressofcsu{Department of Physics, California State University, Dominguez Hills, Carson, CA 90747, USA}\\
\addressofgmu{Department of Physics, George Mason University, Fairfax, VA 22030, USA }\\
\addressofgifu{Department of Physics, Gifu University, Gifu, Gifu 501-1193, Japan}\\
\addressofuh{Department of Physics and Astronomy, University of Hawaii, Honolulu, HI 96822, USA}\\
\addressofkek{Institute of Particle and Nuclear Studies, High Energy Accelerator Research Organization (KEK), Tsukuba, Ibaraki 305-0801, Japan }\\
\addressofkobe{Department of Physics, Kobe University, Kobe, Hyogo 657-8501, Japan}\\
\addressofkyoto{Department of Physics, Kyoto University, Kyoto 606-8502, Japan}\\
\addressoflanl{Physics Division, P-23, Los Alamos National Laboratory, Los Alamos, NM 87544, USA }\\
\addressoflsu{Department of Physics and Astronomy, Louisiana State University, Baton Rouge, LA 70803, USA }\\
\addressofumd{Department of Physics, University of Maryland, College Park, MD 20742, USA }\\
\addressofduluth{Department of Physics, University of Minnesota
Duluth, MN 55812-2496, USA}\\
\addressofsuny{Department of Physics and Astronomy, State University of New York, Stony Brook, NY 11794-3800, USA}\\
\addressofniigata{Department of Physics, Niigata University, Niigata, Niigata 950-2181, Japan }\\
\addressofosaka{Department of Physics, Osaka University, Toyonaka, Osaka 560-0043, Japan}\\
\addressofseoul{Department of Physics, Seoul National University, Seoul 151-742, Korea}\\
\addressofshizuokasc{International and Cultural Studies, Shizuoka Seika College, Yaizu, Shizuoka, 425-8611, Japan}\\
\addressofshizuoka{Department of Systems Engineering, Shizuoka University, Hamamatsu, Shizuoka 432-8561, Japan}\\
\addressoftohoku{Research Center for Neutrino Science, Tohoku University, Sendai, Miyagi 980-8578, Japan}\\
\addressoftokyo{The University of Tokyo, Tokyo 113-0033, Japan }\\
\addressoftokai{Department of Physics, Tokai University, Hiratsuka, Kanagawa 259-1292, Japan}\\
\addressoftit{Department of Physics, Tokyo Institute for Technology, Meguro, Tokyo 152-8551, Japan }\\
\addressofwarsaw{Institute of Experimental Physics, Warsaw University, 00-681 Warsaw, Poland }\\
\addressofuw{Department of Physics, University of Washington, Seattle, WA 98195-1560, USA}\\
}
\affiliation{ } 


\begin{abstract}

Solar neutrino measurements from 1258 days of data from the
Super-Kamiokande detector are presented\footnote{This preprint 
is almost identical to the report submitted to Physical Review Letter.  
We have added to this preprint a few tables of numerical values
that were omitted from the PRL draft.}.  The measurements are based
on recoil electrons in the energy range 5.0--20.0~MeV.  The measured
solar neutrino flux is $2.32 \pm
0.03$~(stat.)~$^{+0.08}_{-0.07}$~(sys.)~$\times
10^6$~cm$^{-2}$~s$^{-1}$, which is
45.1~$\pm$~0.5~(stat.)~$^{+1.6}_{-1.4}$~(sys.)\% of that predicted by
the BP2000 SSM.  The day {\it vs} night flux asymmetry $(\Phi_n -
\Phi_d)/\Phi_{average}$ is $0.033 \pm
0.022$~(stat.)~$^{+0.013}_{-0.012}$~(sys.).  The recoil electron
energy spectrum is consistent with no spectral distortion
($\chi^2/d.o.f. = 19.0/18$).  The seasonal variation of the flux is
consistent with that expected from the eccentricity of the Earth's
orbit ($\chi^2/d.o.f. = 3.7/7$).  For the \hep\ neutrino flux, we set
a 90\% C.L. upper limit of $ 40 \times 10^3$~cm$^{-2}$~s$^{-1}$, which
is 4.3 times the BP2000 SSM prediction.

\end{abstract}


\maketitle


Solar neutrinos have been detected using chlorine-, gallium-, and
water-based
detectors~\cite{homestake-results,kamIII-results,sage-results,gallex-results,sk-flux-300};
all have measured significantly lower solar neutrino flux than
predicted by Standard Solar Models
(SSMs)~\cite{bp2000,bp98,other_ssm}. This disagreement between the
measured and expected solar neutrino flux, known as the ``solar
neutrino problem'', is generally believed to be due to neutrino flavor
oscillations.
Signatures of neutrino oscillations in Super-Kamiokande (SK) might
include distortion of the recoil electron energy (\erecoil) spectrum,
difference between the night-time solar neutrino flux relative to the
day-time flux, or a seasonal variation in the neutrino flux.
Observation of these effects would be strong evidence in support of
solar neutrino oscillations independent of absolute flux calculations.
Conversely, non-observation would constrain oscillation solutions to
the solar neutrino problem.  We describe here solar neutrino
measurements from 1258 days of SK data.

SK, located at Kamioka Observatory, Institute for Cosmic Ray Research,
University of Tokyo, is a 22.5~kton fiducial volume water Cherenkov
detector that detects solar neutrinos via the elastic scattering of
neutrinos off atomic electrons.  The scattered recoil electron is
detected via Cherenkov light production, allowing both the direction
and total energy to be measured.  These quantities are related to the
original neutrino direction and energy.  Detailed descriptions of SK
can be found
elsewhere~\cite{sk-flux-300,linac,sk-espectrum-500,sk-dn-500}.



The 1258-day solar neutrino data were collected in four periods with
different trigger thresholds between May 31, 1996 and October 6, 2000
(table~\ref{tab:sle_numbers}).  The analysis threshold has been at
5.0~MeV except for the first 280 days where the data were analyzed
with a threshold of 6.5~MeV.  The analysis threshold is determined by
the level of irreducible background events and the event trigger
threshold.  An event is triggered when the sum of PMTs registering a
hit in a 200~nsec time window (\nhit) is above a threshold
(table~\ref{tab:sle_numbers}).  This threshold should be sufficiently
low that the trigger efficiency at the analysis threshold is nearly
100\%.  The lowering of the trigger threshold in periods~2--4 was made
possible by the addition of a software filter to the data acquisition
system that removes a large portion of background events.
This removal is accomplished by 
reconstructing the event vertex and rejecting events with vertices
within 2~m of the inner detector wall, most of which are due to
external radioactivity.  Each lowering of the trigger threshold in the
course of the experiment was made possible by increasing the number of
computers that run the filter program.
\begin{table}
\begin{tabular}{c c c c r}
\hline
\hline
  Run    & \nhit     & 50/95\%      & Analysis   &  Live-time  \\
  period & threshold & efficiency   & threshold  &   (days)    \\
         &           & (MeV)        & (MeV)      &             \\ 
\hline
(1) May  $1996 \sim$ & 40.6 & 5.7 / 6.2 & 6.5 &  280 \\
(2) May  $1997 \sim$ & 34.5 & 4.7 / 5.2 & 5.0 &  650 \\
(3) Sep. $1999 \sim$ & 30.4 & 4.2 / 4.6 & 5.0 &  320 \\
(4) Sep. $2000 \sim$ & 27.7 & 3.7 / 4.2 & 5.0 &    8 \\
\hline
\hline
\end{tabular}
\caption{\rm The trigger and analysis thresholds and live-times during
which they were used.  The third column shows the recoil electron
energy at which the trigger is 50\% and 95\% efficient.  The software
filter was added starting in May 1997.}
\label{tab:sle_numbers}
\end{table}


There are $2.0 \times 10^9$ events in the raw data sample before
background reduction.  After removing cosmic ray muon events, the
sample in the 22.5~kton fiducial volume with energy between
5.0--20.0~MeV contains $3.0 \times 10^7$ events.  The dominant
background sources in the low-energy region ($E \lesssim 6.5$~MeV) are
$^{222}$Rn in the water and external radioactivity; in the high-energy
region ($E \gtrsim 6.5$~MeV), radioactive decay of muon-induced
spallation products accounts for most of the background.  Background
reduction takes place in the following steps: first reduction,
spallation cut, second reduction, and external gamma-ray cut.  The
first reduction includes cuts that remove events due to electronic
noise and arcing PMTs.  In addition, a cut on the goodness of the
reconstructed vertex is used to remove obvious background events originating
from various non-physical sources.  
The number of remaining events after the first
reduction is $1.5 \times 10^7$.  The spallation cut has been improved
compared to that used in earlier
publications~\cite{sk-flux-300,sk-espectrum-500,sk-dn-500}. 
We have improved the likelihood functions used in removing spallation
events and introduce a new cut for \nsixteen\ events that originate
from absorption of cosmic ray stopped $\mu^-$ on \osixteen.
The number of events in the high-energy region (6.5--20~MeV) before
and after spallation cut is $1.6 \times 10^6$ and $3.3 \times 10^5$,
respectively.  The spallation cut is 79\% efficient for solar neutrino
events.
The second reduction removes events with poor vertex fit quality or
with blurred Cherenkov ring patterns, characteristics of low-energy
background events and external gamma rays.  This newly introduced
reduction step has improved the signal-to-noise ratio in the
low-energy region by almost an order of magnitude.  The number of
events before and after the second reduction in the 5.0--6.5~MeV
region are $ 1.0 \times 10^7 $ and $ 1.4 \times 10^6 $ events,
respectively.  In addition, the gamma-ray cut, which removes external
events, has been tightened for those events with $E < 6.5$~MeV.  The
combined efficiency of the first reduction, second reduction, and the
external gamma ray cut for solar neutrino events is $\sim 73\%$ for $E
\ge 6.5$~MeV, and $\sim 52$\% for $E < 6.5$~MeV.  After these
reduction steps, 236,140 events remain in the fiducial volume above
5~MeV, with $S/N \approx 1$ in the solar direction.


The SK detector simulation is based on GEANT 3.21~\cite{geant}.  The
energy scale was measured using a larger sample of data from an {\it
in situ} electron linear accelerator~\cite{linac} (LINAC) compared to
that used in earlier results.  The detector simulation's reliability
was tested using the well-known $\beta$ decay of \nsixteen, which is
produced {\it in situ} by an $(n,p)$ reaction on \osixteen.  Fast
neutrons for this reaction are produced using a portable
deuterium-tritium neutron generator (DTG)~\cite{dt-paper}.  The energy
scale measured by the DTG agrees with that from the LINAC within $\pm
0.3$\%.  The total systematic uncertainty in the absolute energy
scale, including possible long term variation and direction
dependence, is $\pm 0.6$\%.


We compare our solar neutrino measurements against reference fluxes
and neutrino spectra in order to search for signatures of neutrino
oscillations.  For $\erecoil \ge 5.0$~MeV, solar neutrinos are
expected to come almost exclusively from the $\beta$ decay of \beight,
with a slight admixture of neutrinos from $^3$He-proton (\hep) fusion.
For the absolute flux of \beight\ and \hep\ neutrinos, we take the
BP2000~\cite{bp2000} SSM as our 
reference.
The $\beta$ decay spectrum of the \beight\ neutrinos is dominated by the
transition to a broad excited state of $^8$Be, which decays
immediately to two $\alpha$ particles.  Bahcall {\it et
al.}~\cite{b8-jnb} use a neutrino spectrum deduced from a comparison
of world data on $^8$Be $\alpha$ decay~\cite{b8-fc,b8-wa,b8-dbw} with
the direct measurement of the positron spectrum from \beight\ decay
measured by Napolitano, Freedman, and Camp~\cite{b8-freedman}.
Energy-dependent systematic errors are deduced from a combination of
experimental uncertainties and the theoretical uncertainties in
radiative and other corrections that must be made to convert the
charged particle data into a neutrino spectrum~\cite{b8-jnb}.
Recently, Ortiz {\it et al.}~\cite{b8-ortiz} have made an improved
measurement of the \beight\ spectrum based on $^8$Be $\alpha$ decay in
which some of the major sources of systematic errors present in
previous measurements were reduced or eliminated.  We have adopted the
neutrino spectral shape and experimental uncertainties from this
measurement.  These experimental uncertainties were then added in
quadrature with the theoretical uncertainties given by Bahcall {\it et
al.}~\cite{b8-jnb}.

\begin{figure}
\includegraphics[width=8cm,clip]{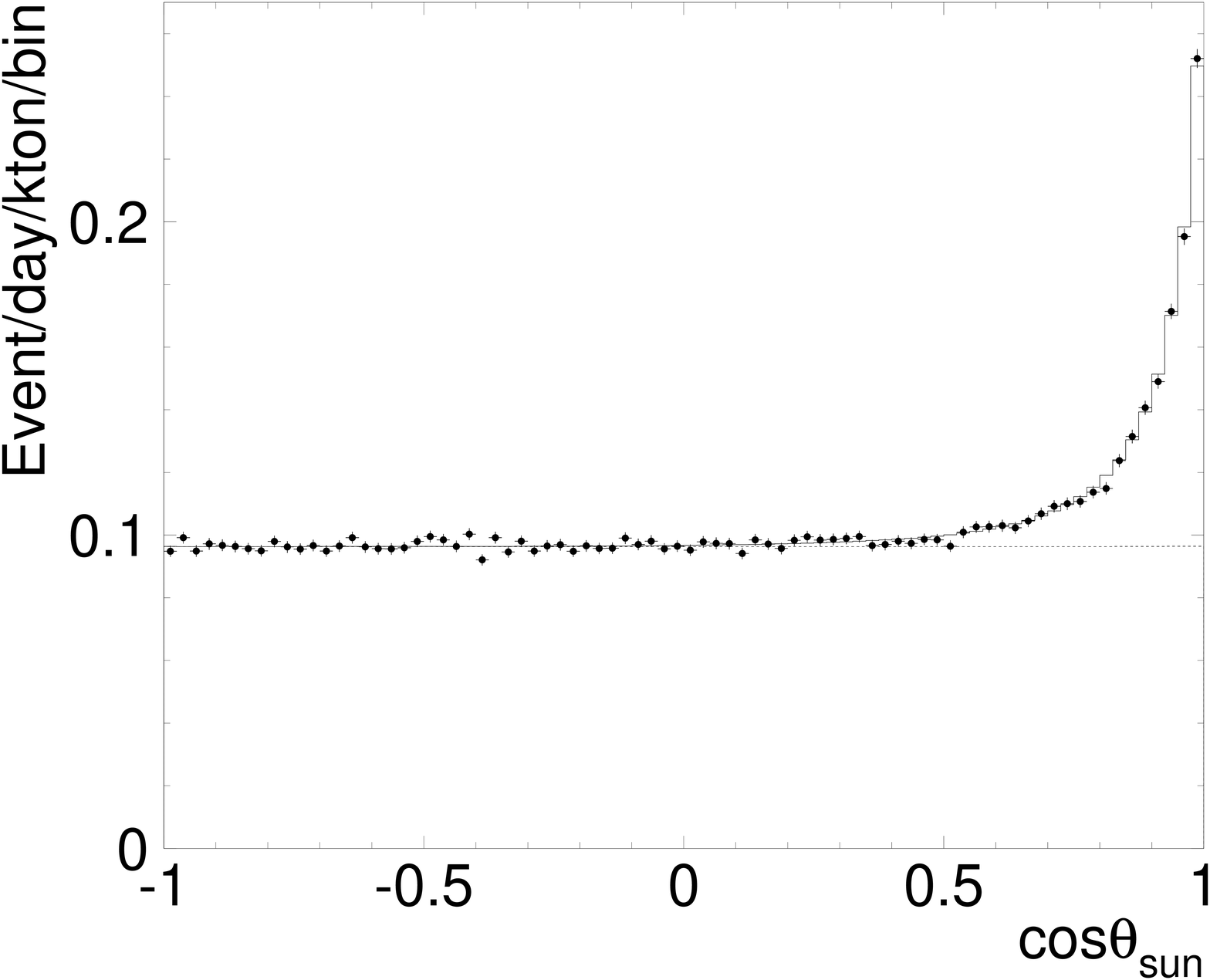}
\caption{\cossun\ distribution for reconstructed energy $E =
$~5.0--20.0~MeV.  The points represent observed data, the histogram
shows the best-fit signal level plus background, and the nearly
horizontal line shows the estimated background.  The peak at
\cossun~=~1 is due to solar neutrinos.}
\label{cossun_a}
\end{figure}
\begin{figure}
\includegraphics[width=8cm,clip]{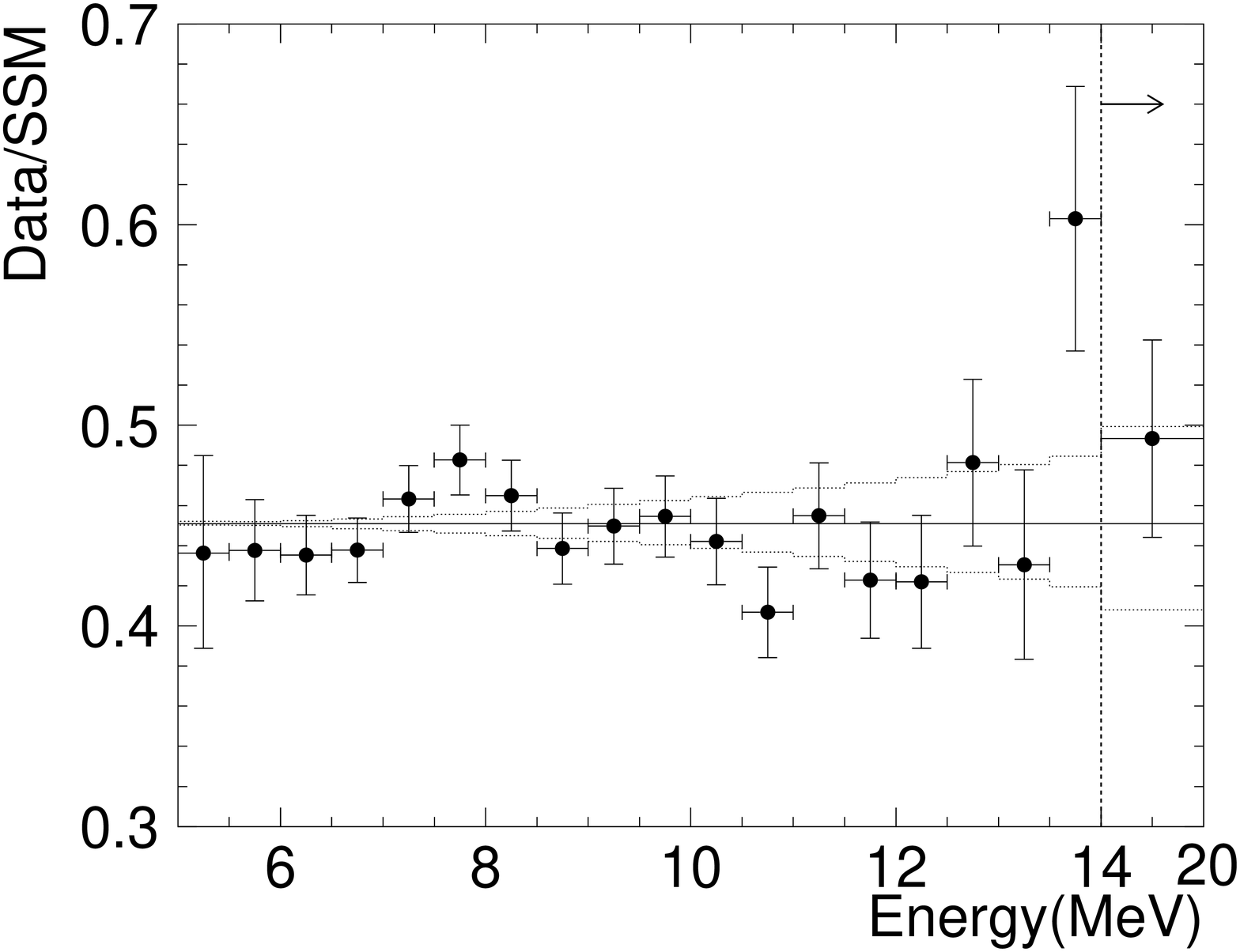}
\caption{The measured \beight\ + \hep\ solar neutrino spectrum
relative to that of Ortiz {\it et al.} normalized to BP2000.  The data
from 14~MeV to 20~MeV are combined into a single bin.  The horizontal
solid line shows the measured total flux, while the dotted band around
this line indicates the energy correlated uncertainty.  Error bars
show statistical and energy-uncorrelated errors added in quadrature.}
\label{spec}
\end{figure}

The solar neutrino signal is extracted from the data using the
\cossun\ distribution (Fig.~\ref{cossun_a}).  
The angle \thetasun\ is that between the recoil electron momentum and
the vector from the sun to the Earth.  The solar neutrino flux is
obtained by a likelihood fit of the signal and background shapes to
the \cossun\ distribution in data.  The signal shape is obtained from
the known angular distribution and detector simulation, while the
background shape is nearly flat in \cossun.  In the \beight\ flux
measurement, the data are subdivided into 19 energy bins in the range
5.0--20.0~MeV (binning as in Fig.~\ref{spec}).  The likelihood
function is defined as follows: {\small
\begin{equation}
{\cal L} =  \prod_{j = 1}^{19} 
  \frac{e^{-(Y_j \cdot S + B_j)}}{N_j!} 
  \prod_{i = 1}^{N_j} \left[ B_j \cdot F_b(\cos\theta_i, E_i)
                       + Y_j \cdot S \cdot F_s(\cos\theta_i, E_i) \right]
\label{eqn:likelihood}
\end{equation}
}
\noindent 
S is the total number of signal events, while $N_j$, $B_j$, and $Y_j$
represent the number of observed events, the number of background
events, and the expected fraction of signal events in the $j$-th bin,
respectively.  $F_b$ and $F_s$ are the probability for the background
and signal events as a function of \cossun\ and energy ($E_i$) of each
event.  The likelihood function is maximized with respect to $S$ and
$B_j$.  For the energy spectrum measurement, each term in the product
over bins is maximized separately.


The best-fit value of $S$ is $18,464 \pm 204~({\rm
stat.})^{+646}_{-554}~({\rm sys.})$, which is
45.1~$\pm$~0.5~(stat.)~$^{+1.6}_{-1.4}$~(sys.)\% of the reference
prediction.  The corresponding \beight\ flux is:
\begin{displaymath}
 2.32 \pm 0.03~({\rm stat.}) ^{+0.08}_{-0.07}~({\rm sys.})
 \times 10^6~{\rm cm}^{-2} {\rm s}^{-1}.
\end{displaymath}
The total systematic error is $^{+3.5\%}_{-3.0\%}$, with the largest
sources coming from the reduction cut efficiency
($^{+2.2\%}_{-1.7\%}$), energy scale and resolution ($\pm 1.4\%$),
systematic shifts in the event vertex ($\pm 1.3\%$), and the angular
resolution of the recoil electron momentum ($\pm 1.2\%$).

\begin{figure}
\includegraphics[width=8cm,clip]{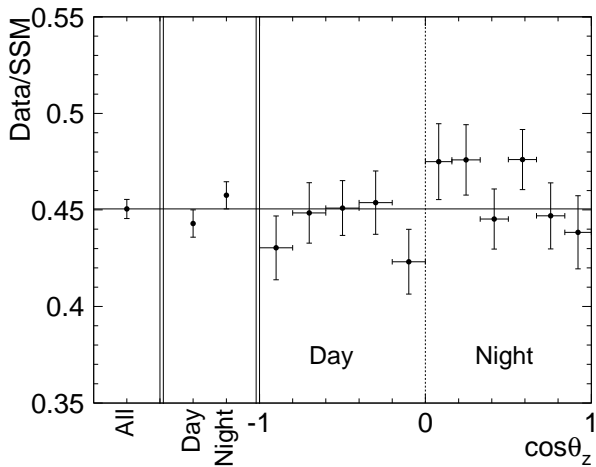}
\caption{The solar zenith angle (\thetaz) dependence of the solar
neutrino flux (error bars show statistical error).  The width of the
night-time bins was chosen to separate solar neutrinos that pass
through the Earth's dense core ($\cosz \ge 0.84$) from those that pass
through the mantle ($0 < \cosz < 0.84$).  
The horizontal line shows the flux for all data.}
\label{dn}
\end{figure}

Fig.~\ref{dn} shows the solar neutrino flux as a function of the solar
zenith angle \thetaz\ (the angle between the vertical axis at SK and
the vector from the sun to the Earth).  
Numerical values are shown in Table~\ref{tab:dn}.
The day-time solar neutrino flux $\Phi_d$ is defined as the flux of events
when $\cosz \leq 0$, while the night-time flux $\Phi_n$ is that when
$\cosz > 0$.  The measured fluxes are:
\begin{eqnarray*}
\Phi_d & = & 2.28 \pm 0.04~({\rm stat.}) ^{+0.08}_{-0.07}~({\rm sys.})
\times 10^6~{\rm cm}^{-2} {\rm s}^{-1} \\
\Phi_n & = & 2.36 \pm 0.04~({\rm stat.}) ^{+0.08}_{-0.07}~({\rm sys.})
\times 10^6~{\rm cm}^{-2} {\rm s}^{-1}
\end{eqnarray*}
Some neutrino oscillation parameters predict a non-zero difference
between $\Phi_n$ and $\Phi_d$ due to the matter effect in the Earth's
mantle and core~\cite{matter-effect}. The degree of this difference is
measured by the day-night asymmetry, defined as ${\cal A} = (\Phi_n -
\Phi_d)/\Phi_{average}$, where $\Phi_{average} = \frac{1}{2} (\Phi_n + 
\Phi_d)$.  We find:
\begin{displaymath}
{\cal A} = 0.033 \pm 0.022~(\rm{stat.})^{+0.013}_{-0.012}~(\rm{sys.})
\end{displaymath}
Including systematic errors, this is $1.3~\sigma$ from zero asymmetry.
Many sources of systematic errors cancel out in the day-night
asymmetry measurement.  The largest sources of error in the asymmetry
are the energy scale and resolution ($^{+0.012}_{-0.011}$) and the
non-flat background shape of the \cossun\ distribution ($\pm 0.004$).

Fig.~\ref{spec} shows the measured recoil electron energy spectrum 
relative to the Ortiz {\it et al.} spectrum normalized to BP2000.
(See Table~\ref{tab:spectrum} for numerical values.)
A fit to an undistorted energy spectrum gives $\chi^2/d.o.f. = 19.0/18$.
Energy-correlated systematic errors are considered in the definition
of $\chi^2$~\cite{sk-espectrum-500}.  The energy-correlated systematic
error (shown in Fig.~\ref{spec} as a band around the total flux) is
due to uncertainties that could cause a systematic shift in the energy
spectrum.  The sources of this error are uncertainties in the energy
scale, resolution, and the reference \beight\ spectrum against which
the data are compared.

\begin{figure}
\includegraphics[width=8cm,clip]{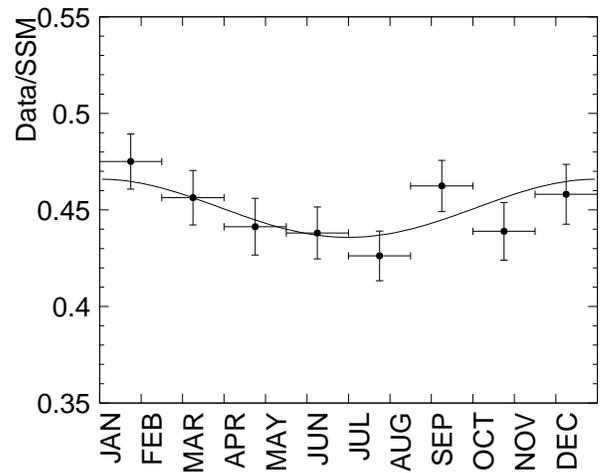}
\caption{Seasonal variation of the solar neutrino flux. The curve
shows the expected seasonal variation of the flux introduced by the
eccentricity of the Earth's orbit.  Error bars show statistical errors
only.}
\label{seasonal}
\end{figure}

The seasonal dependence of the solar neutrino flux is shown in
Fig.~\ref{seasonal}.
(See Table~\ref{tab:seasonal} for numerical values.)
The points represent the measured flux, and the curve shows the
expected variation due to the orbital eccentricity of the Earth
(assuming no neutrino oscillations, and normalized to the measured
total flux).  The data are consistent with the expected annual
variation ($\chi^2/d.o.f. = 3.7/7$).  Systematic errors are included
in the calculation of $\chi^2$.  The total systematic error on the
relative flux values in each seasonal bin is $\pm 1.3\%$, the largest
sources coming from energy scale and resolution ($^{+1.2\%}_{-1.1\%}$)
and reduction cut efficiency ($\pm 0.6\%$).

\begin{figure}
\includegraphics[width=8cm,clip]{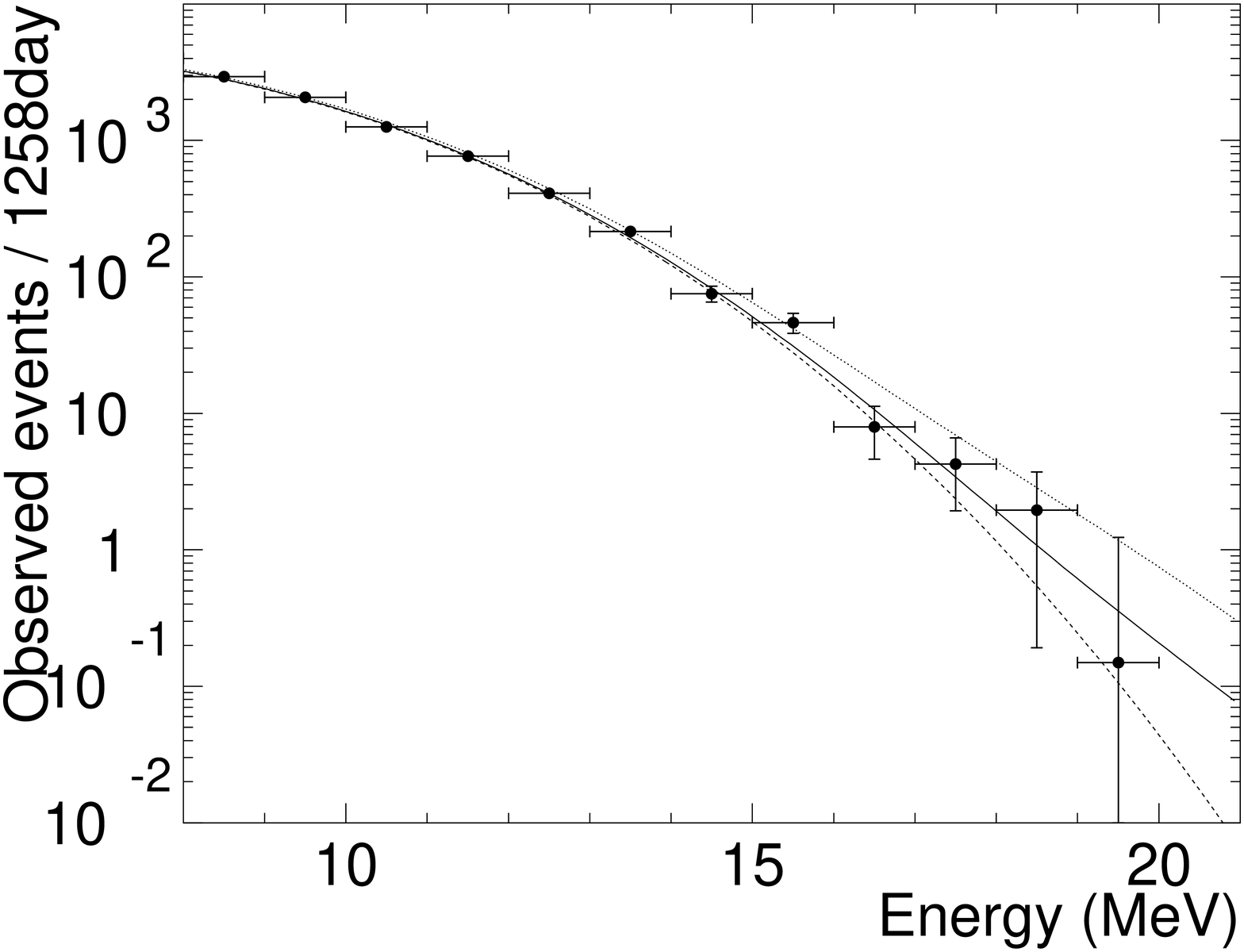}
\caption{Energy spectrum of recoil electrons produced by \beight\ and
\hep\ neutrinos, in 1~MeV bins.  The points show data with statistical
error bars.  The curves show expected spectra with various \hep\
contributions to the best-fit \beight\ spectrum.  The solid, dotted,
and dashed curves show the spectrum with 1, 4.3, and 0 times the
BP2000 \hep\ flux, respectively.}
\label{hep}
\end{figure}

The \hep\ neutrino flux given by BP2000 is $9.3 \times
10^{3}$~cm$^{-2}$~s$^{-1}$~\cite{bp2000}, which is three orders of
magnitude smaller than the \beight\ neutrino flux.  Although the
theoretically calculated \hep\ flux is highly uncertain because of
many delicate cancellations in calculating the astrophysical $S$
factor, a recent calculation by Marcucci, {\it et
al.}~\cite{hep-marcucci} shows that the flux is unlikely to be
drastically larger than the value given in BP2000.  The effect of
\hep\ neutrinos on solar neutrino measurements at SK is expected to be
small.  However, since the end-point of the \hep\ neutrino spectrum is
18.77~MeV compared to about 16~MeV for the \beight\ spectrum, the high
energy end of the \erecoil\ spectrum should be relatively enriched
with \hep\ neutrinos.  An unexpectedly large \hep\ flux may distort
the \erecoil\ spectrum.  In our measurement of the \hep\ flux, we
extract the number of events in the window \erecoil\ = 18--21~MeV from
a \cossun\ distribution like the one shown in Fig.~\ref{cossun_a}.
This window was chosen because it optimizes the significance of the
\hep\ flux measurement in MC assuming BP2000 \beight\ and \hep\
fluxes.  We find $1.3 \pm 2.0$ events in the chosen window.  Assuming
that all of these events are due to \hep\ neutrinos, the 90\%
confidence level upper limit of the \hep\ neutrino flux is $40 \times
10^3$~cm$^{-2}$~s$^{-1}$ (4.3 times the BP2000 prediction).
Fig.~\ref{hep} shows the expected energy spectra with various \hep\
contributions.


In summary, SK has lowered the analysis energy threshold to 5.0~MeV,
collected more than twice the data previously reported, and reduced
systematic errors through refinements in data analysis and extensive
detector calibrations.  With those improvements, and with the 18,464
observed solar neutrino events, SK provides very precise measurements
of the recoil electron energy spectrum, day-night flux asymmetry, and
the absolute solar neutrino flux.  The measured flux is
45.1~$\pm$~0.5~(stat.)~$^{+1.6}_{-1.4}$~(sys.)\% of the BP2000
prediction.  We found no statistically significant energy spectrum
distortion ($\chi^2/d.o.f. = 19.0/18$ relative to the predicted
\beight\ spectrum), and the day-night flux difference of 3.3\% of the
average flux is 1.3~$\sigma$ from zero.  However, the precision of
these measurements should provide strong and important constraints on
the neutrino oscillation parameters.  The seasonal dependence of the
flux shows the expected 7\% annal variation due to the eccentricity of
the Earth's orbit.  This is the first neutrino-based observation of
the Earth's orbital eccentricity.
A stringent limit on the hep neutrino flux ($\Phi_{hep} < 40 \times
10^3$~cm$^{-2}$~s$^{-1}$) was obtained, which corresponds to 4.3 times
the predicted value from BP2000.

The authors acknowledge the cooperation of the Kamioka Mining and
Smelting Company.  The Super-Kamiokande detector has been built and
operated from funding by the Japanese Ministry of Education, Culture,
Sports, Science and Technology, the U.S. Department of Energy, and the
U.S. National Science Foundation.  This work was partially supported
by the Korean Research Foundation (BK21) and the Korea Ministry of
Science and Technology.

\begin{table*}
\begin{center}
\begin{tabular}{|c|c|c|c|}  \hline
Day-Night    & Zenith angle                 & Data/SSM                  & $\delta_i $  \\
\hline \hline
DAY  & $ -1 \leq \cos\theta_{z} \leq 0    $ & $ 0.443^{+0.007}_{-0.007} $ & ------ \\
\hline
MAN1  & $ 0    < \cos\theta_{z} \leq 0.16 $ & $ 0.475^{+0.020}_{-0.020} $ & $^{+1.3}_{-1.2}$\% \\
MAN2  & $ 0.16 < \cos\theta_{z} \leq 0.33 $ & $ 0.476^{+0.018}_{-0.018} $ & $^{+1.3}_{-1.2}$\% \\
MAN3  & $ 0.33 < \cos\theta_{z} \leq 0.50 $ & $ 0.445^{+0.016}_{-0.016} $ & $^{+1.3}_{-1.2}$\% \\
MAN4  & $ 0.50 < \cos\theta_{z} \leq 0.67 $ & $ 0.476^{+0.016}_{-0.016} $ & $^{+1.3}_{-1.2}$\% \\
MAN5  & $ 0.67 < \cos\theta_{z} \leq 0.84 $ & $ 0.447^{+0.017}_{-0.017} $ & $^{+1.3}_{-1.2}$\% \\
CORE  & $ 0.84 < \cos\theta_{z} \leq 1    $ & $ 0.438^{+0.019}_{-0.019} $ & $^{+1.3}_{-1.2}$\% \\
\hline
\end{tabular}
\end{center}
\caption{
Numerical results of the day/night analysis.
The zenith angle region (2nd column),
the ratio of observed and expected number of events (3rd column),
and 1$\sigma$ error of systematic error (4th column).
Systematic errors are relative to Day-flux.
Energy range is 5.0--20.0MeV.
}
\label{tab:dn}
\end{table*}
\begin{table*}
\begin{center}
\begin{tabular}{|c|c|c|c|c|c|}  \hline
Energy    & \multicolumn{3}{c|}{Data/SSM} & $\delta_{i,cor.}$ & $\delta_{i,uncor.}$ \\
\hline
(MeV)     &  ALL & DAY & NIGHT & &  \\ \hline \hline
5.0-5.5   & $ 0.436^{+0.046}_{-0.046} $ & $0.438^{+0.065}_{-0.065}$ & $0.434^{+0.063}_{-0.063}$ & $^{+0.2}_{-0.2}$\% & $^{+3.9}_{-3.1}$\% \\
5.5-6.0   & $ 0.438^{+0.024}_{-0.024} $ & $0.428^{+0.034}_{-0.034}$ & $0.446^{+0.034}_{-0.034}$ & $^{+0.2}_{-0.2}$\% & $^{+1.7}_{-1.6}$\% \\
6.0-6.5   & $ 0.435^{+0.019}_{-0.019} $ & $0.426^{+0.027}_{-0.027}$ & $0.444^{+0.027}_{-0.027}$ & $^{+0.3}_{-0.3}$\% & $^{+1.3}_{-1.4}$\% \\
6.5-7.0   & $ 0.438^{+0.015}_{-0.015} $ & $0.431^{+0.021}_{-0.021}$ & $0.444^{+0.021}_{-0.021}$ & $^{+0.5}_{-0.6}$\% & $^{+1.4}_{-1.4}$\% \\
7.0-7.5   & $ 0.463^{+0.015}_{-0.015} $ & $0.462^{+0.022}_{-0.022}$ & $0.464^{+0.022}_{-0.022}$ & $^{+0.8}_{-0.8}$\% & $^{+1.3}_{-1.4}$\% \\
7.5-8.0   & $ 0.483^{+0.016}_{-0.016} $ & $0.494^{+0.023}_{-0.023}$ & $0.472^{+0.022}_{-0.022}$ & $^{+1.0}_{-1.1}$\% & $^{+1.3}_{-1.4}$\% \\
8.0-8.5   & $ 0.465^{+0.017}_{-0.017} $ & $0.452^{+0.023}_{-0.023}$ & $0.477^{+0.023}_{-0.023}$ & $^{+1.4}_{-1.3}$\% & $^{+1.3}_{-1.4}$\% \\
8.5-9.0   & $ 0.438^{+0.017}_{-0.017} $ & $0.402^{+0.024}_{-0.024}$ & $0.473^{+0.024}_{-0.024}$ & $^{+1.7}_{-1.7}$\% & $^{+1.3}_{-1.4}$\% \\
9.0-9.5   & $ 0.450^{+0.018}_{-0.018} $ & $0.454^{+0.026}_{-0.026}$ & $0.446^{+0.025}_{-0.025}$ & $^{+2.1}_{-2.0}$\% & $^{+1.3}_{-1.4}$\% \\
9.5-10.0  & $ 0.455^{+0.019}_{-0.019} $ & $0.449^{+0.027}_{-0.027}$ & $0.460^{+0.027}_{-0.027}$ & $^{+2.5}_{-2.3}$\% & $^{+1.3}_{-1.4}$\% \\
10.0-10.5 & $ 0.442^{+0.021}_{-0.021} $ & $0.430^{+0.029}_{-0.029}$ & $0.454^{+0.029}_{-0.029}$ & $^{+3.0}_{-2.7}$\% & $^{+1.3}_{-1.4}$\% \\
10.5-11.0 & $ 0.407^{+0.022}_{-0.022} $ & $0.386^{+0.030}_{-0.030}$ & $0.426^{+0.032}_{-0.032}$ & $^{+3.4}_{-3.2}$\% & $^{+1.3}_{-1.4}$\% \\
11.0-11.5 & $ 0.455^{+0.026}_{-0.026} $ & $0.439^{+0.036}_{-0.036}$ & $0.470^{+0.037}_{-0.037}$ & $^{+3.9}_{-3.6}$\% & $^{+1.3}_{-1.4}$\% \\
11.5-12.0 & $ 0.423^{+0.028}_{-0.028} $ & $0.455^{+0.042}_{-0.042}$ & $0.394^{+0.038}_{-0.038}$ & $^{+4.5}_{-4.2}$\% & $^{+1.3}_{-1.4}$\% \\
12.0-12.5 & $ 0.422^{+0.033}_{-0.033} $ & $0.389^{+0.045}_{-0.045}$ & $0.455^{+0.047}_{-0.047}$ & $^{+5.1}_{-4.8}$\% & $^{+1.3}_{-1.4}$\% \\
12.5-13.0 & $ 0.481^{+0.041}_{-0.041} $ & $0.514^{+0.061}_{-0.061}$ & $0.451^{+0.055}_{-0.055}$ & $^{+5.8}_{-5.4}$\% & $^{+1.3}_{-1.4}$\% \\
13.0-13.5 & $ 0.431^{+0.047}_{-0.047} $ & $0.468^{+0.070}_{-0.070}$ & $0.397^{+0.063}_{-0.063}$ & $^{+6.5}_{-6.2}$\% & $^{+1.3}_{-1.4}$\% \\
13.5-14.0 & $ 0.603^{+0.065}_{-0.065} $ & $0.551^{+0.092}_{-0.092}$ & $0.653^{+0.094}_{-0.094}$ & $^{+7.4}_{-7.0}$\% & $^{+1.3}_{-1.4}$\% \\
14.0-20.0 & $ 0.493^{+0.049}_{-0.049} $ & $0.430^{+0.067}_{-0.067}$ & $0.559^{+0.071}_{-0.071}$ & $^{+10.7}_{-9.5}$\% & $^{+1.3}_{-1.4}$\% \\
\hline
\end{tabular}
\end{center}
\caption{
Numerical results of the energy spectrum analysis.
The ratio of observed and expected number of events in all-time (2nd column),
in day-time (3rd column),
in night-time (4th column),
1$\sigma$ error of correlated systematic error (5th column),
and 1$\sigma$ error of uncorrelated systematic error (6th column).
Systematic errors are relative.
}
\label{tab:spectrum}
\end{table*}
\begin{table*}
\begin{center}
\begin{tabular}{|c|c|c|c|}  \hline
Time period &   Data/SSM  &    Data/SSM at 1 AU & $\delta_i $  \\
\hline \hline
Jan.  1 -- Feb. 16  & $ 0.475^{+0.014}_{-0.014} $ & $ 0.461^{+0.014}_{-0.014} $ & $^{+1.3}_{-1.3}$\% \\
Feb. 17 -- Apr.  2  & $ 0.456^{+0.014}_{-0.014} $ & $ 0.450^{+0.014}_{-0.014} $ & $^{+1.3}_{-1.3}$\% \\
Apr.  3 -- May  19  & $ 0.441^{+0.015}_{-0.015} $ & $ 0.446^{+0.015}_{-0.015} $ & $^{+1.3}_{-1.3}$\% \\
May  20 -- Jul.  4  & $ 0.438^{+0.013}_{-0.013} $ & $ 0.451^{+0.014}_{-0.014} $ & $^{+1.3}_{-1.3}$\% \\
Jul.  5 -- Aug. 19  & $ 0.426^{+0.013}_{-0.013} $ & $ 0.439^{+0.013}_{-0.013} $ & $^{+1.3}_{-1.3}$\% \\
Aug. 20 -- Oct.  4  & $ 0.462^{+0.013}_{-0.013} $ & $ 0.469^{+0.014}_{-0.014} $ & $^{+1.3}_{-1.3}$\% \\
Oct.  5 -- Nov. 18  & $ 0.439^{+0.015}_{-0.015} $ & $ 0.434^{+0.015}_{-0.015} $ & $^{+1.3}_{-1.3}$\% \\
Nov. 19 -- Dec. 31  & $ 0.458^{+0.016}_{-0.016} $ & $ 0.445^{+0.015}_{-0.015} $ & $^{+1.3}_{-1.3}$\% \\
\hline
\end{tabular}
\end{center}
\caption{
Numerical results of the seasonal analysis.
The time period (1st column),
the ratio of observed and expected number of events without eccentricity correction (2nd column),
the ratio of observed and expected number of events with eccentricity correction (3rd column),
and 1$\sigma$ error of systematic error (4th column).
Systematic errors are relative.
Energy range is 5.0--20.0MeV.
}
\label{tab:seasonal}
\end{table*}

\end{document}